\documentclass[aps,prl,showpacs,amsfonts,amssymb,floatfix,twocolumn,floats]{revtex4}
\usepackage{graphicx}
\usepackage{bm}
\begin{document}

\title{Low-temperature nucleation in a kinetic Ising model with soft 
stochastic dynamics}
\author{Kyungwha Park$^{1,2}$}
\author{Per Arne Rikvold$^{2,3}$}
\author{Gloria M.\ Buend{\'\i}a$^{2,3,4}$}
\author{M.~A.\ Novotny$^{3}$}
\affiliation{
$^1$Center for Computational Materials Science,
Code 6390, Naval Research Laboratory, Washington D.C. 20375\\
$^2$School of Computational Science and Information Technology,
Center for Materials Research and Technology, and Department of Physics,
Florida State University, Tallahassee, Florida 32306-4120\\
$^3$Department of Physics and Astronomy and ERC Center for Computational
Sciences, Mississippi State University, Mississippi State,
Mississippi 39762-5167\\
$^4$Department of Physics, Universidad
Sim{\'o}n Bol{\'\i}var, Caracas 1080, Venezuela
}
\date{\today}
\begin{abstract}
We study low-temperature nucleation in kinetic Ising models by
analytical and simulational methods, confirming the
general result for the average metastable lifetime, 
$\langle \tau \rangle = A \exp (\beta \Gamma )$ ($\beta = 1/k_{\rm B} T$) 
[E.\ Jord{\~a}o Neves and R.~H. Schonmann, Commun.\ Math.\ Phys.\ {\bf
137}, 209 (1991)]. Contrary to common belief, 
we find that both $A$ and  $\Gamma$ depend 
significantly on the stochastic dynamic. 
In particular, for a ``soft'' dynamic, in which the effects of 
the interactions and the applied field
factorize in the transition rates, $\Gamma$ does {\it not\/} simply equal the
energy barrier against nucleation, as it does for 
the standard Glauber dynamic, which does not have this factorization property.
\end{abstract}


\maketitle

Nucleation is fundamentally important in 
disciplines ranging from biochemistry \cite{ONUC} to earth sciences
\cite{LASA}, astrophysics \cite{PATZ}, and cosmology \cite{KAST98},
and it has been studied by kinetic Monte Carlo (MC) simulations in 
electrochemistry \cite{BROW99}, materials science 
\cite{COMB00}, 
magnetism \cite{NOVO02A}, and atmospheric science \cite{MAHN03}, 
to mention just a few. 
However, many questions in nucleation theory are still 
unresolved, and recently
there has been much interest in kinetic Ising systems as models 
for nucleation.  In particular, much work has been done on their
dynamical behavior at very low temperatures 
\cite{NEVE91,NOVO02,PARK02,BOVI02,SHNE02,SHNE03},
where it is influenced by lattice discreteness. It is then
possible to calculate exactly both the shape of the 
critical nucleus (the saddle-point configuration)
and the most probable path during a nucleation event. 
In a typical numerical experiment, the system is prepared in a metastable
state with all spins positive in a negative applied field. During each 
MC step (MCS), a randomly chosen spin is flipped with a
configuration-dependent transition rate $W$ that satisfies detailed
balance, so that it will drive the system to thermodynamic equilibrium.
The metastable lifetime is measured as the average number of
MCS until the magnetization reaches zero. 
In the regime of single-droplet decay 
studied here, the lifetime measured in MCS is independent of the system size
\cite{NEVE91,RIKV94A,PARK02}. 
In the low-temperature limit the lifetime has been 
rigorously shown to be \cite{NEVE91} 
\begin{equation}
\langle \tau \rangle = A e^{\beta \Gamma} .
\label{eq:tau}
\end{equation}
Here the only dependence on the temperature $T$ is through 
$\beta = 1/ k_{\rm B} T$, where $k_{\rm
B}$ is Boltzmann's constant (hereafter set equal to one).
It is often assumed that $\Gamma$ equals the energy difference  
between the saddle point and the metastable state \cite{SHNE02,SHNE03},
independent of the specific stochastic dynamic. 
In this Letter we show that this is not
always so. In particular, we describe two dynamics that both obey
detailed balance 
but have different values of $\Gamma$ and $A$ for all values of the
applied field, despite having the same saddle-point configuration. 

At sufficiently low $T$, the saddle point was shown in Ref.~\cite{NEVE91} to
be an $\ell \times ( \ell - 1)$ rectangle of overturned spins  
with a ``knob'' of one overturned spin on one of its 
long sides. The critical length  
$\ell = \lfloor 2 J / |H| \rfloor + 1$ 
for all $|H| \in (0,4)$, 
where $\lfloor x \rfloor$ is the integer part of $x$. Here, $J>0$ is the 
nearest-neighbor interaction constant of the Ising model, which will
henceforth be set to unity. The critical length thus
changes discontinuously at values of $|H|$ such that $2 / |H|$ is an integer. 

The square-lattice $S=1/2$ Ising ferromagnet with unit interaction 
is defined by the Hamiltonian 
${\cal H} = - \sum_{\langle \alpha,\beta \rangle} \sigma_\alpha \sigma_\beta 
- H \sum_{\alpha} \sigma_\alpha$, 
where the Ising spins $\sigma_\alpha = \pm 1$, 
$H$ is the applied field, $\sum_{\langle \alpha,\beta \rangle}$ runs
over all nearest-neighbor bonds on a square lattice, and $\sum_{\alpha}$ runs
over all lattice sites. When this system evolves under a
continuous-time Glauber dynamic with spin-flip rate \cite{GLAU63}
\begin{equation}
W_{\mathrm G} = [1+\exp{(\beta \Delta E)}]^{-1}
\;,
\label{eq:G}
\end{equation}
where $\Delta E$ is the energy change that would result from the
flip, $\Gamma$ in Eq.~(\ref{eq:tau}) is given by 
\cite{NEVE91}
\begin{equation}
\Gamma_{\rm Hard} = 8 \ell - 2 |H| ( \ell^2 - \ell + 1 ) \;,
\label{eq:gamma}
\end{equation}
and from Ref.~\cite{BOVI02}
$A = A_{\rm Hard} = 3/[8 ( \ell - 1)]$ for all $|H| < 2$. 
(See explanation of the subscript ``Hard'' below.)
The interpretation of $\Gamma_{\rm Hard}$ is indeed the
energy difference between the saddle point and the metastable state. 

A characteristic feature of the Glauber dynamic is that it does not
factorize into one part that depends only on the change in interaction
energy, $\Delta E_J$, and another that depends only on the change in the
field energy, $\Delta E_H$. 
Such transition rates are known as ``hard'' \cite{MARR99}. 
\begin{widetext}

\begin{table}[h]
\caption[]{
Rates of flipping a spin in class $m$, $p_m$, in the limit 
$\beta \rightarrow \infty$. Here $\sigma = +$ ($-$) corresponds to a spin 
in the metastable (stable) direction, and $N_+$ is
the number of its nearest-neighbor spins in the metastable direction.
The analytic form of $p_m$ for the soft dynamic does not change 
with $|H|$.
} 
\label{tab:prob}
\begin{tabular}{c|c|c|c|c||c|c|c|c|c}
$m$ & $\sigma$ & $N_+$ & $p_m^{\rm Soft}$   & $p_m^{\rm Hard}$
& $m$ & $\sigma$ & $N_+$ & $p_m^{\rm Soft}$ & $p_m^{\rm Hard}$  \\ \hline
1   & +   & 4     & $e^{-\beta 8}$        & $e^{-\beta (8-2|H|)}$ for $|H|< 4$
& 6 & $-$ & 4     & $e^{-\beta 2|H|}$     & 1 for $|H|< 4$  \\ 
    &     &       &                       & 1/2 for $|H|=4$ 
&   &     &       &                       & 1/2 for $|H|=4$ \\  
    &     &       &                       & 1 for $|H|>4$
&   &     &       &                       & $e^{\beta (8-2|H|)}$ for $|H|> 4$  
                                                                      \\ \hline 
2   & +   & 3     & $e^{-\beta 4}$        & $e^{-\beta (4-2|H|)}$ for $|H|< 2$
& 7 & $-$ & 3     & $e^{-\beta 2|H|}$     & 1 for $|H|< 2$ \\
    &     &       &                       & 1/2 for $|H|=2$ 
&   &     &       &                       & 1/2 for $|H|=2$ \\
    &     &       &                       & 1 for $|H|>2$
&   &     &       &                       & $e^{\beta (4-2|H|)}$ for $|H|> 2$ 
                                                                      \\ \hline
3   & +   & 2     & 1/2                   & 
1 for $|H|>0$
& 8 & $-$ & 2     & $e^{-\beta 2|H|}$/2   & $e^{- \beta 2 |H|}$ for all $|H|$ 
                                                                      \\ \hline
4   & +   & 1     & 1                     & 1 for all $|H|$ 
& 9 & $-$ & 1     & $e^{-\beta (4+2|H|)}$ & $e^{-\beta (4+2|H|)}$ for all $|H|$  
                                                                      \\ \hline
5   & +   & 0     & 1                      & 1 for all $|H|$
&10 & $-$ & 0     & $e^{-\beta (8+2|H|)}$ & $e^{-\beta (8+2|H|)}$ for all $|H|$ 
                                                                      \\ 
\end{tabular}
\end{table}

\end{widetext}
Dynamics that do factorize this way are
called ``soft.'' An example is the soft Glauber dynamic \cite{RIKV02},
\begin{equation}
W_{\mathrm{SG}} = 
[1+\exp{(\beta \Delta E_J)}]^{-1} \,
[1+\exp{(\beta \Delta E_H)}]^{-1}
\;.
\label{eq:SG}
\end{equation}

In studies of field-driven Ising and solid-on-solid interfaces 
\cite{RIKV02,RIKV00B} 
it was recently shown  
that soft dynamics yield significantly different microscopic interface
structures and mobilities than hard dynamics. Here we show
that also the low-temperature nucleation properties with the 
soft Glauber dynamic 
differ significantly from those with the hard Glauber dynamic. 
In particular, $\Gamma$ is not
simply the energy difference between the saddle point and the metastable state, and the prefactor $A$ is also different. 

We obtain our results in three different ways. First, we calculate 
analytically by hand the first-passage time from the metastable state
to an absorbing state just beyond the saddle point 
in an approximation that the path in configuration
space corresponds to a simple one-step Markov process \cite{KAMP92}. 
Second, we perform computer-aided analytical calculations  
using the technique of absorbing
Markov chains (AMC) \cite{NOVO02,IOSI80}, allowing for 
multiple branching paths and ``blind alleys.'' 
Third, we perform simulations using the MC with AMC
(MCAMC) technique \cite{NOVO02,NOVO95}.
The first method provides the clearest physical insight, and for noninteger 
values of $2/|H|$ the results are fully confirmed by the other two. 

\begin{figure}[t]
\includegraphics[angle=0,width=.43\textwidth]{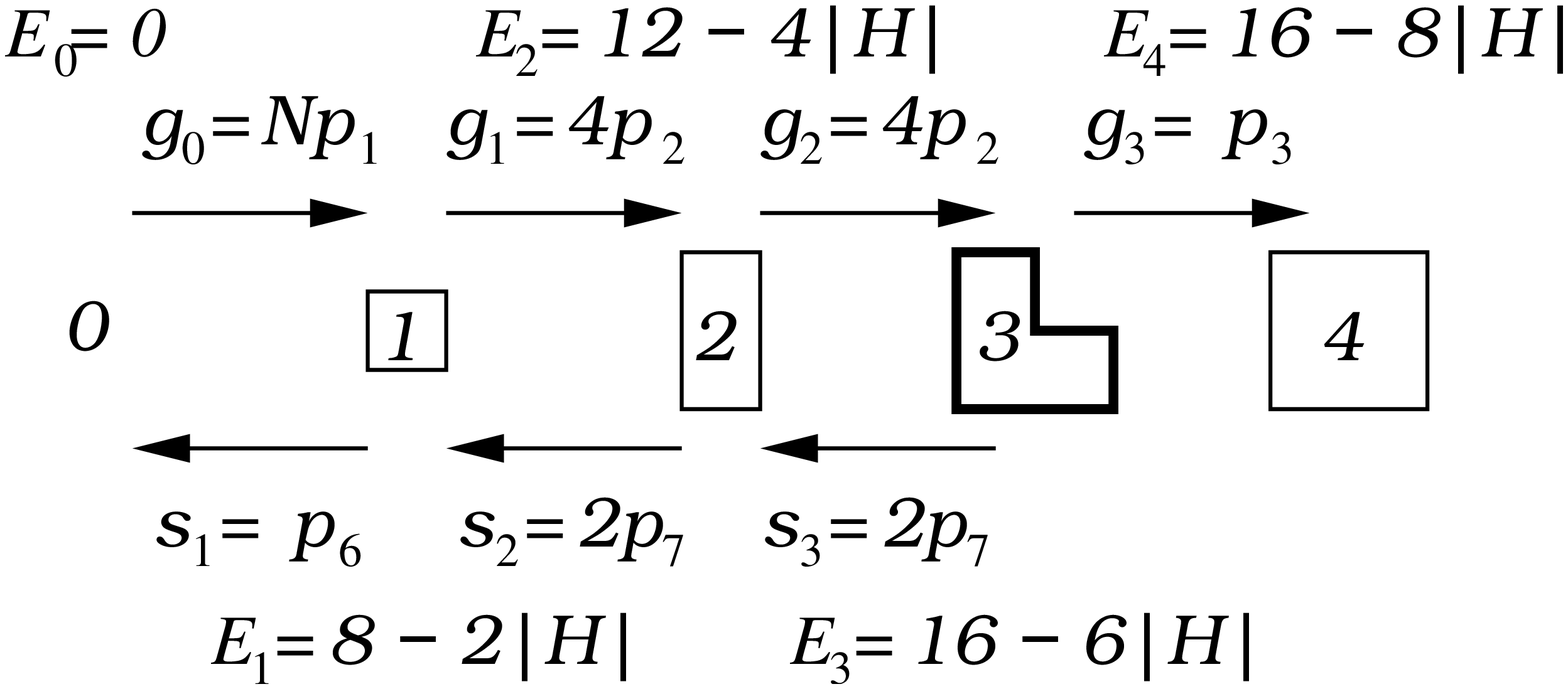}
\vspace{.3cm}
\caption[]{
The states in the one-step 
Markov chain of clusters of $i = 0,...,4$ overturned spins, 
used to calculate the metastable lifetime $\langle \tau \rangle$ 
analytically by hand. The right-pointing 
arrows give the growth rates $g_{i-1}$, and the left-pointing arrows give the 
shrinkage probabilities $s_i$ for $i =1$, 2, and~3. 
The energies $E_i$ (relative to 
the metastable state, $i=0$) are given at the top of the figure for even $i$ 
and at the bottom for odd $i$. 
}
\label{fig:conf}
\end{figure}

The one-step Markov chain for $1 < |H| < 2$ ($\ell = 2$) 
corresponds to the configurations labeled $i = 0,...,4$ in
Fig.~\ref{fig:conf}. The label $i$ gives the number of overturned 
spins, such that the starting configuration has
$i=0$, and the saddle point has $i = i^* =3$. In general the
absorbing state is labeled $I \ge i^* \ge 1$. The mean 
time spent in state $i$ is $h_i$. The rate at which the cluster grows
from $i$ to $i+1$ overturned spins is $g_i$, 
and the rate with which it shrinks from $i$ to $i-1$ is $s_i$. 
(Multiple spin flips are negligibly rare in the  
zero-temperature limit \cite{NEVE91}.)
These quantities satisfy the relation \cite{KAMP92,KOLE98B,NOVO99B}
\begin{equation}
h_{i-1} = (s_i h_i + N )/ g_{i-1}
\label{eq:Miro}
\end{equation}
with boundary conditions $s_I = s_0 = 0$. 
The number $N$ of sites in the system represents the total
probability current through the Markov chain \cite{KAMP92}. From 
Eq.~(\ref{eq:Miro}) we obtain $h_i$ recursively as $h_{I-1} = N/g_{I-1}$ and 
\begin{equation}
h_i = \frac{N}{g_i} + \sum_{k=1}^{I-1-i} \frac{N}{g_{i+k}} 
                      \prod_{j=1}^k \left( \frac{s_{i+j}}{g_{i+j-1}} \right)
\label{eq:Miro2}
\end{equation}
for $0 \le i \le I-2$. Assuming that $I > i^*$ and the growth 
time for the supercritical droplet is negligible compared with the 
nucleation time \cite{NEVE91,RIKV02,RIKV00B}, 
the lifetime $\langle \tau \rangle$ is the mean first-passage time to $I$, 
$\langle \tau \rangle = \langle \tau_I \rangle = \sum_{i=0}^{I-1} h_i$. 
Grouping the terms according to the ``unpaired'' factors
$N/g_{i+k}$ in Eq.~(\ref{eq:Miro2}) then yields
\begin{equation}
\langle \tau_I \rangle 
=
\frac{N}{g_0} + \sum_{l=1}^{I-1} \frac{N}{g_l} 
\left( 1 + \sum_{k=1}^l 
\prod_{j=0}^{k-1} \frac{s_{l-j}}{g_{l-j-1}} 
\right) 
\;.
\label{eq:Miro3}
\end{equation}
This result is general for any one-step Markov chain with absorption at
$I$, regardless of the values of $g_i$ and $s_i$ \cite{KAMP92}. 
However, for the Ising
model the transition rates are related by detailed balance as
$
( {s_i / n^s_i}) / ( { g_{i-1}  / n^g_{i-1}} ) = 
e^{\beta (E_i - E_{i-1})}$,
where $E_i$ is the energy of state $i$. 
The degeneracy factors $n^s_i$ and $n^g_{i-1}$ are the
numbers of lattice sites at which a single spin flip can shrink the
cluster from $i$ to $i-1$ and analogously for growth
from $i-1$ to $i$, respectively. As a result, Eq.~(\ref{eq:Miro3}) becomes 
\begin{equation}
\langle \tau_I \rangle 
=
\frac{N}{g_0} + \sum_{l=1}^{I-1} \frac{N}{g_l} 
\left( 1 + \sum_{k=1}^l 
e^{\beta (E_l - E_{l-k})} 
\prod_{j=0}^{k-1} \frac{n^s_{l-j}}{n^g_{l-j-1}} 
\right) 
\;.
\label{eq:Miro4}
\end{equation}
In the limit $\beta \rightarrow \infty$ Eq.~(\ref{eq:Miro4}) is dominated
by the term or terms with the largest exponential factor. 
Their selection, which determines $\Gamma$ and $A$,
is described below, after we next find the spin-flip rates 
in the different dynamics. 

For the square-lattice Ising system, the spins fall
into 10 classes \cite{NOVO95}, determined by the spin value $\sigma$
($+$ for the metastable direction and $-$ for the stable direction)
and the number $N_+$ of its nearest neighbors that point in the metastable
direction. The low-temperature limits of the rates $p_m$ for flipping
a spin in class $m$ [Eqs.~(\ref{eq:G}) and~(\ref{eq:SG})],
are shown in Table~\ref{tab:prob}.

Figure~\ref{fig:conf} shows a one-step
Markov chain with $I=4$. For $1 < |H| < 2$ the 
saddle-point configuration ($i=i^*$) 
is the L-shaped cluster with $\ell = 2$ ($i^*=3$).
Among the dominant terms in Eq.~(\ref{eq:Miro4})
is always the term with $k = l = i^* -1$. 
For $0 < |H| < 2$, growth from $i^*-1$ 
to $i^*$ always involves adding a ``knob'' to 
one of the long sides of an $\ell \times (\ell -1)$ 
rectangle, such that $g_{i^* -1} = 2 \ell p_2$.
From Table~\ref{tab:prob} we see that 
$p_2^{\rm Soft} = e^{- \beta 2 |H|} p_2^{\rm Hard}$. 
For $2<|H|<4$, the saddle point is a single overturned spin 
($\ell = 1$), so $i^* =1$ and $g_{i^* -1} = g_0 = N p_1$. 
Again, the difference between $\Gamma$ for the two
dynamics is determined by the fact that 
$p_1^{\rm Soft} = e^{- \beta 2 |H|} p_1^{\rm Hard}$. This yields our main
result:
\begin{equation}
\Gamma_{\rm Soft} = \Gamma_{\rm Hard} + 2 |H|  \;\;\; {\rm for} \; 0< |H|<4
\;.
\label{eq:gams}
\end{equation}
We emphasize that Eq.~(\ref{eq:gams}) is valid as $T \rightarrow 0$ 
for all $|H| \in (0,4)$,
although the low-temperature regime will only
be be reached at exceedingly low $T$ as $|H|$ decreases. 
For $|H|>4$ the lifetime is the first-passage time to one
overturned spin, so that $\langle \tau \rangle = \langle \tau_1 \rangle 
= 1/p_1$, which yields 
$\Gamma_{\rm Soft} = 8$ and $\Gamma_{\rm Hard} = 0$. Thus, in contrast
to the hard dynamic, 
nucleation with the soft dynamic is always activated, even 
for infinitely strong fields. 

To obtain the prefactors $A_{\rm Soft}$ and $A_{\rm Hard}$ in the two
dynamics, we explicitly write out the four terms obtained from
Eq.~(\ref{eq:Miro4}) 
for $I=4$ with $s_i$, $g_{i-1}$, and $E_i$ from Fig.~\ref{fig:conf}:
\begin{eqnarray}
\langle \tau_4 \rangle 
&=&
{1 \over p_1}  
+ {1 \over 4 p_2 } \left( N + e^{\beta(8 - 2|H|)} \right)          \nonumber\\
&&+ {1 \over 4 p_2 } \left( N + {N \over 2} e^{\beta(4 - 2|H|)} 
                            + {1 \over 2} e^{\beta(12 - 4|H|)} \right)
                                                                   \nonumber\\
&&+ {1 \over p_3 } \left( N + {N \over 2} e^{\beta(4 - 2|H|)} 
                                                             \right. 
                                                                   \nonumber\\
&&                 
                    \left.  
                            + {N \over 4} e^{\beta(8 - 4|H|)} 
                            + {1 \over 4} e^{\beta(16 - 6|H|)} 
			    \right)                                \nonumber\\
&\equiv& {\cal A} + {\cal B} + {\cal C} + {\cal D} 
\;.
\label{eq:tauI4}
\end{eqnarray}
Using $p_m$ from Table~\ref{tab:prob} we identify the dominant 
terms in $\langle \tau_4 \rangle$, and from these we obtain $A$
and $\Gamma$ for both the soft and hard Glauber dynamics for all $|H|>1$. 
(Analogous calculations can be carried out for arbitrarily small $|H|$.)

\begin{figure}[b]
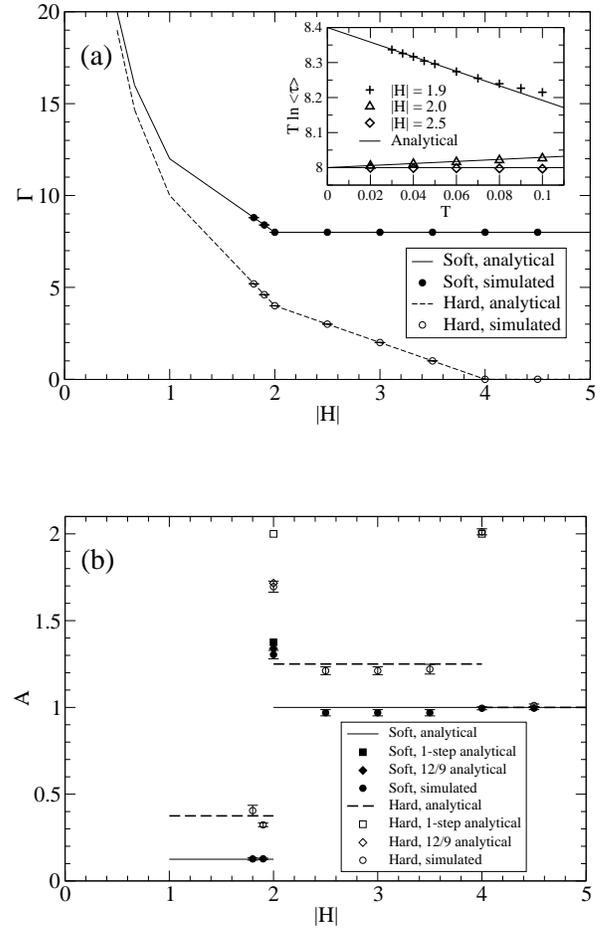

\vspace{0.3truecm}
\includegraphics[angle=0,width=.43\textwidth]{FigGamIns.eps} \\
\vspace{1.2truecm}
\includegraphics[angle=0,width=.43\textwidth]{FigA.eps}
\vspace{0.3truecm}
\caption[]{
Analytical and simulated results for the soft and hard Glauber dynamics
for (a) $\Gamma$ and (b) $A$. In the legends, ``1-step analytical''
refers to the one-step Markov-chain approximation, ``12/9 analytical'' to
the computer-aided AMC calculations with 12 transient and 9 absorbing
configurations, and ``analytical'' to results that
are identical for all the analytical calculations. 
The results only differ for $|H|=2$.
The inset in (a) shows the analytical (lines) and MC (data points)
results for $T \ln \langle \tau \rangle$ vs $T$, from which $\Gamma$ and
$A$ are obtained.
}
\label{fig:GamA}
\end{figure}

Soft dynamic:
For $1 < |H| < 2$ the sum is dominated by the last term in $\cal C$, 
yielding $A_{\rm Soft} = 1/8$ and $\Gamma_{\rm Soft} = 16 - 4|H|$.  
For $|H| = 2$ it is dominated by $\cal A$ and the last terms in 
$\cal B$ and $\cal C$, 
yielding $A_{\rm Soft} = 11/8$ and $\Gamma_{\rm Soft} = 8$.  
For $|H| > 2$ it is dominated by $\cal A$, 
yielding $A_{\rm Soft} = 1$ and $\Gamma_{\rm Soft} = 8$.  
See Fig.~\ref{fig:GamA}. 

Hard dynamic:
For $1 < |H| < 2$ the sum is dominated by the last terms in $\cal C$ and
$\cal D$, 
yielding $A_{\rm Hard} = 3/8$ and $\Gamma_{\rm Hard} = 16 - 6|H|$.  
For $|H| = 2$ it is dominated by $\cal A$ and the last terms in 
$\cal B$, $\cal C$, and $\cal D$, 
yielding $A_{\rm Hard} = 2$ and $\Gamma_{\rm Hard} = 4$.  
For $2 < |H| < 4$ it is dominated by $\cal A$ and the last term in
$\cal B$, 
yielding $A_{\rm Hard} = 5/4$ and $\Gamma_{\rm Hard} = 8 - 2|H|$.  
(These results agree with corresponding ones  in Refs.~\cite{SHNE02,SHNE03}.)
For $|H| \ge 4$ the system is unstable and 
$\langle \tau \rangle = \langle \tau_1 \rangle = {\cal A}$, 
yielding 
$\Gamma_{\rm Hard} = 0$ and  $A_{\rm Hard} = 2$ for $|H|=4$ and 
$A_{\rm Hard} = 1$ for $|H|>4$. 
See Fig.~\ref{fig:GamA}. 

We further 
performed computer-aided analytic calculations of $\langle \tau \rangle$ 
with Mathematica \cite{WOLF96} by the 
AMC method \cite{NOVO02,IOSI80}, using three different classifications
of the configurations: 12 transient and 9 absorbing states (denoted 
12/9), as well as 7/13 and 13/13. For noninteger $2/|H|$ the
results were {\it identical\/} to the 1-step approximation. However, for
$|H|=2$, $A$ (but not $\Gamma$) was found to depend slightly on the numbers of 
states included in the calculation for both dynamics. With the
numbers of states used, these differences were less than 0.5\%. 
Specifically, 12/9 yielded 
$A_{\rm Hard} = 78244/45597 \approx 1.7160$ 
(1.764 by a different method in Ref.~\cite{SHNE02})
and 
$A_{\rm Soft} = 943/704 \approx 1.3395$. 
See Fig.~\ref{fig:GamA}.  

Both sets of analytic results were checked by 
MC simulations for both dynamics,
using the MCAMC method \cite{NOVO02,NOVO95}. 
The system size was $L=24$, and 2000 escapes were used
(6000 for $|H| \ge 4$). The parameters $\Gamma$ and $A$ were determined
from weighted two-parameter linear least-squares fits to plots of 
$T \ln \langle \tau \rangle$ vs $T$ [inset in Fig.~\ref{fig:GamA}(a)]. 
As seen in Fig.~\ref{fig:GamA}, 
the simulation results agree with the analytical
results to within two standard errors. 

In conclusion, we have confirmed Eq.~(\ref{eq:tau}) \cite{NEVE91} for the
low-temperature metastable lifetime of a kinetic Ising model, using both
analytical methods and MC simulations, finding both
$\Gamma$ and $A$ to depend on the specific stochastic
dynamic for all values of the applied field. 
For a soft Glauber dynamic [Eq.~(\ref{eq:SG})],
$\Gamma$ does {\it not\/} equal the energy
difference between the critical cluster and the metastable state 
for any value of the field and it also does {\it not\/} 
vanish in the strong-field
limit, as it does for the conventional, hard Glauber dynamic
[Eq.~(\ref{eq:G})]. Thus, nucleation under the soft dynamic remains
an activated process for arbitrarily strong fields.
These results are consistent with recent studies
of the microstructure and mobility 
of field-driven Ising and Solid-on-Solid interfaces \cite{RIKV02,RIKV00B}.
They indicate that great caution must be shown in formulating and
interpreting stochastic models of physical systems, as even seemingly
minor modifications of the transition probabilities can significantly
affect the nucleation rates. 
It might thus be interesting to investigate the
influence of the specific stochastic dynamic on dynamic phase
transitions in kinetic Ising models \cite{SIDE98}. 
We also note that, although our results are
derived for a specific model system, qualitatively similar results
should apply to kinetic MC simulations for  
nucleation in a wide range of scientific disciplines. 
On the positive side, experimental observation of the field
and temperature dependences of nucleation and growth
could help devise correct stochastic models of nonequilibrium phenomena.

\begin{acknowledgments}
We gratefully acknowledge correspondence with V.~Shneidman. 
P.A.R.\ and G.M.B.\ appreciate the
hospitality of the Mississippi State University
Department of Physics and Astronomy and ERC Center for Computational Sciences.
This work was supported by NSF grants No.\ DMR-0120310 and DMR-0240078.
\end{acknowledgments}



\end{document}